# Unfolding of vortices into topological stripes in a multiferroic material

X. Wang[1], M. Mostovoy[2], M. G. Han[3], Y. Horibe[1], T. Aoki[4], Y. Zhu[3], and S.-W. Cheong[1,*]

[1] Rutgers Center for Emergent Materials and Department of Physics and Astronomy, Rutgers University, Piscataway, NJ 08854, USA

[2] Zernile Institute for Advanced Materials, University of Groningen, Nijenborgh 4, 9747 AG, Groningen, The Netherlands

[3] Condensed Matter Physics and Materials Science, Brookhaven National Laboratory, Upton, NY 11973, USA

[4] JEOL USA, Inc., Peabody, MA 01960, USA

*Corresponding author: sangc@physics.rutgers.edu

**Multiferroic hexagonal $RMnO_3$ (R=rare earths) crystals exhibit dense networks of vortex lines at which six domain walls merge. While the domain walls can be readily moved with an applied electric field, the vortex cores were so far impossible to control. Our experiments demonstrate that shear strain induces a Magnus-type force pulling vortices and antivortices in opposite directions and unfolding them into a topological stripe domain state. We discuss the analogy between this effect and the current-driven dynamics of vortices in superconductors and superfluids.**

Understanding and controlling topological defects [1] in ordered states with spontaneously broken symmetries is essential for technological applications of functional materials. For example, ferromagnetic and ferroelectric memory devices rely on field- and current-induced dynamics of domain walls. Unusual physical properties of topological defects open new ways for their manipulation. The electric polarization induced by chiral magnetic domain walls, for example, was used to shift them with an inhomogeneous electric field [2]. Vortices in superfluid liquids and superconductors drift under the influence of the Magnus force in a direction normal to both the supercurrent and vorticity vectors [3, 4]. The transverse force also acts on skyrmions flowing with the electrical current in helimagnetic conductors [5]. Here we show that shear strain has a similar effect on topological defects in multiferroic hexagonal manganites: it moves vortices and antivortices in opposite directions and stabilizes a topological stripe domain state, in which all domain walls have the same chirality.

Hexagonal (h-) $RMnO_3$ (R=rare earths) are improper ferroelectrics below the transition into a trimerized lattice state caused by the size mismatch between R and Mn-O layers[6-9]. The domains in h-$RMnO_3$ associated with two polarization directions and three trimerization phases can have two topologically distinct configurations: vortices [10-12] and stripes [13] (see Fig. 1(a) and (b)). Stripes formed by domain walls spanned over the entire sample are extremely robust, whereas the density of vortices strongly varies depending on heat treatment conditions. When h-$RMnO_3$ crystals are grown below the ferroelectric-trimerization transition temperature, $T_c$, which can reach as high as 1450 °C in the case of $LuMnO_3$, they exhibit stripe domains of large widths. However, when a h-$RMnO_3$ crystal with stripe domains is heated above and cooled down across $T_c$, vortex domains form everywhere in the crystal. The density of vortices decreases with the decreasing cooling rate across $T_c$ Slow relaxational dynamics should eventually expel all vortices from the crystal, leaving possibly stripe-type domains [12]. However, unaided by external stimuli, this process takes an astronomically large time, and thus the transformation of vortices into stripes has never been directly observed. Topological characteristics of vortex domains, such as vorticity and $Z_2 \times Z_3$ symmetry, have been discussed in literature [14, 15]. At the same time, the topological nature of stripe domain states and the direct relationship between stripes and vortices have not been explored so far. Here, we report the direct observation of the vortex-to-stripe transformation in the presence of shear strain with a gradient, unveiling the topological nature of stripes. We show that strain is an effective tool for vortex control: previously only domain walls could be moved with an applied electric field, while the vortex cores remained unaffected [15-17].

Plate-like hexagonal ErMnO$_3$ single crystals (~1×1×0.02 mm$^3$, thin along the c axis) grown using a Bi$_2$O$_3$-based flux method were used for our experiments. Shear strain was applied on ErMnO$_3$ crystals (EMO-A, B, and C) by putting a weight (alumina rod) on the crystals during a thermal treatment (see Supplementary Information section 1). Crystals of ErMnO$_3$ (T$_c$=1130 °C) were slowly cooled across the ferroelectric-trimerization transition temperature; stayed at 1140 °C for 10mins, and then slowly cooled to 1100 °C with the rate of 30 °C/h, followed by furnace cooling. In order to visualize domain patterns, crystals were etched chemically in phosphoric acid at 150 °C. The domain patterns of chemically etched crystals were observed using optical and atomic force microscopies, as phosphoric acid preferentially etches the surface of upward-polarization domains. A specimen (EMO-D) for high-angle annular-dark-field scanning transmission electron microscopy (HAADF-STEM) imaging with the detection angles ranging from 68 mrad to 280 mrad was prepared using a focused ion beam lift-out technique on a chemical-etched crystal showing the surface domain structure on the *ab* plane. HAADF-STEM images were obtained using a JEOL ARM 200 CF equipped with a CEOS Cs-corrector.

Figure 1(c) shows a low-magnification optical microscope (OM) image of the surface of an ErMnO$_3$ crystal (EMO-A) after applying strain at high temperatures. White dashed lines indicate the position of the alumina rod exerting a downward force on EMO-A. The yellow dashed lines show the location of the edges of the groove in the alumina plate (Supplementary Information section 1). The tilted dark line between the two white lines indicates where the alumina rod touched the crystal. The rod was off-centered in the vertical direction to make the force in the bottom triangular region larger than in the top region. Stripe-like domains along the alumina rod direction (perpendicular to the top edge of EMO-A) were observed near the alumina rod, whereas the remaining area showed vortex domains. The vortex-to-stripe transformation takes place near the boundary where vortices meet stripes, as shown in Fig. 1(d) displaying a high-magnification OM image of the green-boxed area in Fig. 1(c). Vortices evidently were unfolded and became stripes. The opposite surface of EMO-A exhibits similar domain patterns (Supplementary Information section 2).

In order to clarify the mechanism for the vortex-to-stripe transformation, we performed experiments on two more crystals: the rectangular-shaped $ErMnO_3$ (EMO-B) with a centered alumina rod (Fig. 1(e)) and another triangular-shaped $ErMnO_3$ (EMO-C) with an off-centered rod (Supplementary Information section 3). Surprisingly, EMO-B showed only vortices with no hint of stripes, even in the region right under the alumina rod [see Fig. 1(f)], while EMO-C exhibited the vortex-to-stripe transformation, similar to that observed in EMO-A. Evidently, the crystal shape is crucial for the vortex-to-stripe transformation. Figure 1(g) shows schematically the in-plane strain distribution in triangular-shape and rectangular-shape specimens to illustrate what may be happening in those different-shape crystals (the out-of-plane strain is not shown, as it does not couple to the trimerization phase). In rectangular EMO-B, the top surface is compressed under the weight of the alumina rod, while the bottom surface is stretched. The top compressive and the bottom tensile strains cancel each other on average, i.e. in the middle of the crystal (See page 12 in Supplementary Information section 6). The triangular corner of EMO-A and EMO-C exposed to an additional strain. To amplify this effect, we intentionally shifted the center of mass of the alumina rod closer to the triangular corner, which produced an additional shear strain with a large in-plane gradient in the corner (see Fig. 1(g) and Supplementary Information section 4 for estimates of the strain magnitude, which includes Refs. [18, 19]). This average shear strain with a large gradient induces the vortex-to-stripe transformation, as discussed below.

Atomic force microscope (AFM) images, which exhibit higher spatial resolution than OM images, unveil the details of the vortex-to-stripe transformation. The AFM image of EMO-A in Fig. 2(a) corresponds to the optical image in Fig. 1(d). The green-boxed area in Fig. 2(a) is expanded in Fig. 2(b), and the detailed analysis of the purple-boxed area in Fig. 2(b) is shown in Fig. 2(c). There exist three trimerization domains (α, β, and γ) in h-RMnO3, each supporting two directions of electric polarization. All these six domains meet to form a cloverleaf arrangement with alternating polarization direction. There are two types of cloverleaf defects with opposite sequences of domains along a loop encircling a defect, which can be viewed as vortices and antivortices. If we call the defect at the very top of Fig. 2(c) a vortex with the sequence of the trimerized phases (α+, β-, γ+, α-, β+, γ-) along a contour in the clockwise direction (red circle), then all cloverleaf defects in Fig. 2(c) can be labeled in a consistent manner. The resulting vortices and antivortices are marked with red and blue circles, respectively. The dark (bright) regions in AFM images display valleys (mountain plateaus) corresponding to domains with upward (downward) ferroelectric polarization [12-14].

The above analysis reveals several remarkable features: [i] the fixed sequence of the six trimerized phases (…,α+, β-, γ+, α-, β+, γ-, α+, β-,…) in the stripe region from right to left, [ii] the predominant presence of vortices at the transformation boundary (i.e. the applied strain mostly expels antivortices from the sample) and [iii] no vortices or antivortices in the stripe domain region (see Fig. 2(b) and a larger area shown in Fig. 2(a)).

The single chirality of stripes is consistent with the concept put forward by Artyukhin *et al*.[20] that an applied shear strain can stabilize a "Φ–staircase" state due to the interaction

$$F_{int} = -\lambda \int dV \left[ (u_{xx} - u_{yy})\partial_x \Phi - 2u_{xy}\partial_y \Phi \right], \tag{1}$$

where $u_{ij}$ is the strain tensor and Φ is the phase of the commensurate lattice modulation (trimerization). Equation (1) is the so-called Lifshitz invariant discussed, in particular, in the context of commensurate-incommensurate transitions[21]. In the uniform ground state Φ =

$0, \pm\frac{\pi}{3}, \pm\frac{2\pi}{3}$ or $\pi$, corresponding to the six degenerate ferroelectric-trimerization states. Shear strain, on the other hand, favors an incommensurate lattice modulation. The compromise is the stripe state – a periodic array of parallel domain walls with $\Phi$ monotonically increasing or decreasing by $\pi/3$ at each wall (see a complete proof in Supplementary Information section 5).

Figure 3(a) is a schematic for Fig. 2(c), and Fig. 3(b) is a topologically deformed cartoon of transformation boundary shown in Fig 3(a). Consider now the change of $\Phi$ along a closed yellow contour around the boundary between the topological stripe and the vortex-antivortex domains, as depicted in Fig. 3(b). According to Stoke's lemma, $\oint d\boldsymbol{x}\cdot\nabla\Phi = 2\pi(N_V - N_A)$, where $N_V$ and $N_A$ are the total numbers of vortices and antivortices inside the contour, respectively. $\Phi$ changes monotonically only in the line interval $AB$, while in the rest of contour $\Phi$ changes randomly. Thus, for a sufficiently large integration contour, $\Phi_B - \Phi_A = \frac{\pi}{3} N_{DW} = 2\pi(N_V - N_A)$, $N_{DW}$ being the total number of domain walls in the stripe domain. The excess of vortices is, therefore, related to the number of stripes: one extra vortex per 6 domain walls.

The transformation of a random network of topological defects into the array of parallel stripes involves the separation of vortices from antivortices. A uniform shear strain $u_{xx} - u_{yy}$ favoring stripes along the y-axis applies a force to a domain wall connecting a vortex with an antivortex. Using Eq.(1), we find that, independent from the shape of the wall, its total interaction energy is $F_{int} = -\frac{\pi\lambda h}{3}(u_{xx} - u_{yy})(y_V - y_A)$, where $y_V(y_A)$ is the y-coordinate of the vortex(antivortex) and $h$ is the sample thickness. Differentiating this energy with respect to $y_V$ and $y_A$, we obtain the forces that pull the vortex and antivortex away from each other and lead to their unfolding into stripes. This process is illustrated in Figs. 3(c) and 3(d) showing a vortex-antivortex pair. The phase $\Phi$ increases in the positive x-direction in the region between the vortex and antivortex, which in the presence of the strain $u_{xx} - u_{yy}$ lowers $F_{int}$, whereas above the vortex and below the antivortex the phase gradient has the energetically unfavorable direction. The force pulling the vortex and antivortex apart, $F_y = 2\pi\lambda h(u_{xx} - u_{yy})$, increases the length of the energetically favorable domain walls at the expense of the unfavorable ones.

One can draw an analogy between the topological stripe state and a superfluid liquid with a superfluid current proportional to the gradient of the condensate phase. The force on vortices and antivortices in h-RMnO$_3$ generated by strain is analogous to the Magnus force that moves superfluid vortices and antivortices in the direction transverse to the superfluid current (for more discussion, see Supplementary Information section 6, which includes Refs. [22, 23]).

If the applied uniform strain pushes all antivortices out of the sample and all vortices towards the transformation boundary, the presence of a much larger excess of vortices is be expected at the boundary, which is not consistent with our observations. Therefore, the transformation of vortices into stripes must involve an additional mechanism, which may be the coupling between vortices/antivortices and the strain gradient due to the lattice distortion created at the vortex/antivortex core by six merging domain walls. The resulting force pushes vortices and antivortices in the same direction (Supplementary Information section 6). This strain-gradient-induced force can expel extra vortices and antivortices from the strained area to the outside of the specimen. This mechanism is consistent with the fact that the stripes are only observed in the regions where the strain is non-uniform.

The monochirality of stripes is confirmed by our atomic-resolution HAADF-STEM images. A scanning electron microscope (SEM) image of the top surface of EMO-D after chemical etching is shown in Fig. 4(a): narrow valleys are upward (+) polarization domains. A specimen for HAADF-STEM experiments was obtained cutting out the green-box area in Fig. 4(a) with FIB milling, and a dark-field TEM image of the specimen taken with the $00\bar{4}$ reflection near the [100] projection shown in Fig. 4(b) displays the cross-sectional view of domain structures, which are consistent with the SEM image in Fig. 4(a). Atomic-scale HAADF-STEM experiments were performed on the domain walls labeled DW1-DW4, and false-colored HAADF-STEM images of Er-ion columns are shown in Fig. 4(c) (the full HAADF-STEM images are shown and discussed in Supplementary Information section 7). Kumagai and Spaldin have discussed four different types of domain walls in h-RMnO$_3$, which we label A-D: the A and B types have undistorted R ions at the walls, and the C and D types do not have undistorted R ions at the walls [24] (see Supplementary Information section 8 for the schematics of the A-D type domain walls). The results of our HAADF-STEM experiments lead to two important findings: stripe domains are indeed monochiral (i.e. [α-, β+, γ-, α+, β-] along one direction), and domain walls alternate between the A and C types. Although the sampling region in atomic-scale HAADF-STEM experiments is rather small, these results are fully consistent with our results discussed earlier. According to Ref. 18, the lowest energy wall is of C type. The adjacent parallel walls must then be of either A type or D type. Consistently, we find the alternation of A and C type walls. This result suggests that the domain walls around a vortex core probably exhibit the alternation of B and C types (Supplementary Information section 9).

In summary, employing shear strain with a large gradient, we were able to transform vortices and antivortices into stripes in hexagonal ErMnO$_3$. This stripe domain state is monochiral and shows the alternation of walls with distorted and undistorted Er ions. Our findings reveal an unexpected topological relation between vortex and stripe domains in h-RMnO$_3$, and also provide a novel way of controlling topological defects. The mechanical control of topological defects in multiferroics opens a novel route to technological applications including mechanical sensors, transducers and memories.

Acknowledgement:
We thank Seung Chul Chae and Lijun Wu for fruitful discussions, and Hyeon Na Woo (RISD) for providing a three-dimensional cartoon. The work at Rutgers was supported by the National Science Foundation DMR-1104484, and the work at Brookhaven National Laboratory was supported by the U.S. Department of Energy's Office of Basic Energy Science, Division of Materials Science and Engineering DE-AC02-98CH10886.

**References**

[1]   N. D. Mermin, Reviews of Modern Physics **51**, 591 (1979).
[2]   A. S. Logginov, G. A. Meshkov, A. V. Nikolaev, E. P. Nikolaeva, A. P. Pyatakov, and A. K. Zvezdin, Appl. Phys. Lett. **93** (2008).
[3]   E. B. Sonin, Phys. Rev. B **55**, 485 (1997).
[4]   J. Bardeen and M. J. Stephen, Physical Review **140**, A1197 (1965).
[5]   J. D. Zang, M. Mostovoy, J. H. Han, and N. Nagaosa, Phys. Rev. Lett. **107** (2011).
[6]   E. F. Bertaut, F. Forrat, and F. P.H, C.R.Acad. Sci. Paris **256**, 1958 (1963).
[7]   T. Katsufuji, S. Mori, M. Masaki, Y. Moritomo, N. Yamamoto, and H. Takagi, Phys. Rev. B **64**, 104419 (2001).
[8]   C. J. Fennie and K. M. Rabe, Phys. Rev. B **72** (2005).
[9]   B. B. Van Aken, T. T. M. Palstra, A. Filippetti, and N. A. Spaldin, Nat. Mater. **3**, 164 (2004).
[10] T. Jungk, A. Hoffmann, M. Fiebig, and E. Soergel, Appl. Phys. Lett. **97** (2010).
[11] S. M. Griffin, M. Lilienblum, K. T. Delaney, Y. Kumagai, M. Fiebig, and N. A. Spaldin, Phys. Rev. X **2** (2012).
[12] T. Choi, Y. Horibe, H. T. Yi, Y. J. Choi, W. D. Wu, and S. W. Cheong, Nat. Mater. **9**, 253 (2010).
[13] S. C. Chae, N. Lee, Y. Horibe, M. Tanimura, S. Mori, B. Gao, S. Carr, and S. W. Cheong, Phys. Rev. Lett. **108** (2012).
[14] S. C. Chae, Y. Horibe, D. Y. Jeong, S. Rodan, N. Lee, and S. W. Cheong, Proc. Natl. Acad. Sci. U. S. A. **107**, 21366 (2010).
[15] S. C. Chae, Y. Horibe, D. Y. Jeong, N. Lee, K. Iida, M. Tanimura, and S. W. Cheong, Phys. Rev. Lett. **110**, 167601 (2013).
[16] M.-G. Han, Y. Zhu, L. Wu, T. Aoki, V. Volkov, X. Wang, S. C. Chae, Y. S. Oh, and S.-W. Cheong, Adv. Mater. **25**, 2415 (2013).
[17] D. Meier, J. Seidel, A. Cano, K. Delaney, Y. Kumagai, M. Mostovoy, N. A. Spaldin, R. Ramesh, and M. Fiebig, Nat. Mater. **11**, 284 (2012).
[18] M. Poirier, F. Laliberté, L. Pinsard-Gaudart, and A. Revcolevschi, Phys. Rev. B **76**, 174426 (2007).
[19] C. L. Lin, J. Liu, X. D. Li, Y. C. Li, S. Q. Chu, L. Xiong, and R. Li, J. Appl. Phys. **112**, 6, 113512 (2012).
[20] S. Artyukhin, K. T. Delaney, N. A. Spaldin, and M. Mostovoy, Nat. Mater. **13**, 42 (2013).
[21] A. P. Levanyuk, S. A. Minyukov, and A. Cano, Phys. Rev. B **66**, 014111 (2002).
[22] A. Cano, A. P. Levanyuk, and S. A. Minyukov, Phys. Rev. B **68**, 144515 (2003).
[23] P. Miranović, L. Dobrosavljević-Grujić, and V. G. Kogan, Phys. Rev. B **52**, 12852 (1995).
[24] Y. Kumagai and N. A. Spaldin, Nature communications **4**, 1540 (2013).

**Figure Legends**

**FIG. 1** (color online) Effect of the crystal shape for annealing under strain (triangle vs. rectangle). (a) and (b) are OM images of chemically etched $ErMnO_3$ crystals, which indicate two distinct domain patterns: stripes and vortices. (c), Collaged optical microscope image of EMO-A after chemical etching. (d), Enlarged OM image of the green-box area in Fig. 1(c) showing the vortex-to-stripe transformation. (e), OM images of triangular EMO-B. (f), Enlarged image of the green-boxed area in Fig. 1(e) shows only vortices. (g), Schematics of in-plane strain on the top surface, average (or middle region), and bottom surface. Blue, red, and black colors indicate compressive, tensile, and no strain, respectively.

**FIG. 2** (color online) (a), Large-range AFM image of the area in Fig. 1(d) showing vortex-to-stripe transformation. (b), Fine-scan AFM image of the green-boxed area in Fig. 3(a). (c), Expanded AFM image of the purple-box area in Fig. 3(b) with the self-consistently assigned trimerization and ferroelectric phases of all vortex and stripe domains. An excess of vortices (red circles, the clockwise (α-, β+, γ-, α+, β-, γ+) sequence) at the vortex-to-stripe transformation boundary results from predominant expulsion of antivortices (dashed blue circles, the anti-clockwise (α-, β+, γ-, α+, β-, γ+) sequence). All domain walls in stripes exhibit the same chirality corresponding to the sequence (α-, β+, γ-, α+, β-, γ+) from right to left in the whole stripe.

**FIG. 3** (color online) (a) A schematic of Fig. 2(c). (b), A topologically deformed cartoon of the boundary between the vortex-antivortex domains (upper part) and the topological stripe domains (lower part). Vortices (antivortices) are marked with red (blue) circles. (c), A vortex-antivortex pair under the shear $u_{xx} - u_{yy}$ strain inducing vertical forces (light blue arrows) on the vortex and antivortex. Purple arrows indicate the direction of the phase gradient. (d), Vertical stripe domains which are about to form, as the Magnus-type force pulls the vortex and antivortex apart.

**FIG. 4** (color online) STEM-HAADF analysis of different types of domain walls. (a), SEM image of EMO-D ab-plane surface after chemical etching. Mountain plateaus (valleys) are ferroelectric domains with downward (upward) polarization. The green box indicates the area cut for a STEM-HAADF experiments using FIB. (b), Dark-field TEM image of a STEM-HAADF specimen with the [00$\bar{4}$] reflection near the [100] zone axis. Dark lines are narrow ferroelectric domains with upward polarization. Also shown are crystallographic axes and an enlarged image of the red box area. (c), False-colored STEM-HAADF images of 1-4 domain walls in the inset of Fig. 4(b). DW1 and DW3 domain walls are of C type (with distorted Er ions at walls), and DW2 and DW4 domain walls are of A type (with un-distorted Er ions at walls). Each STEM-HAADF image is followed by schematics of Er-ion columns. C-type domain walls are slightly slanted (by a unit cell), so near the wall centers two upward and downward Er columns overlap, giving rise to elongated Er column images.

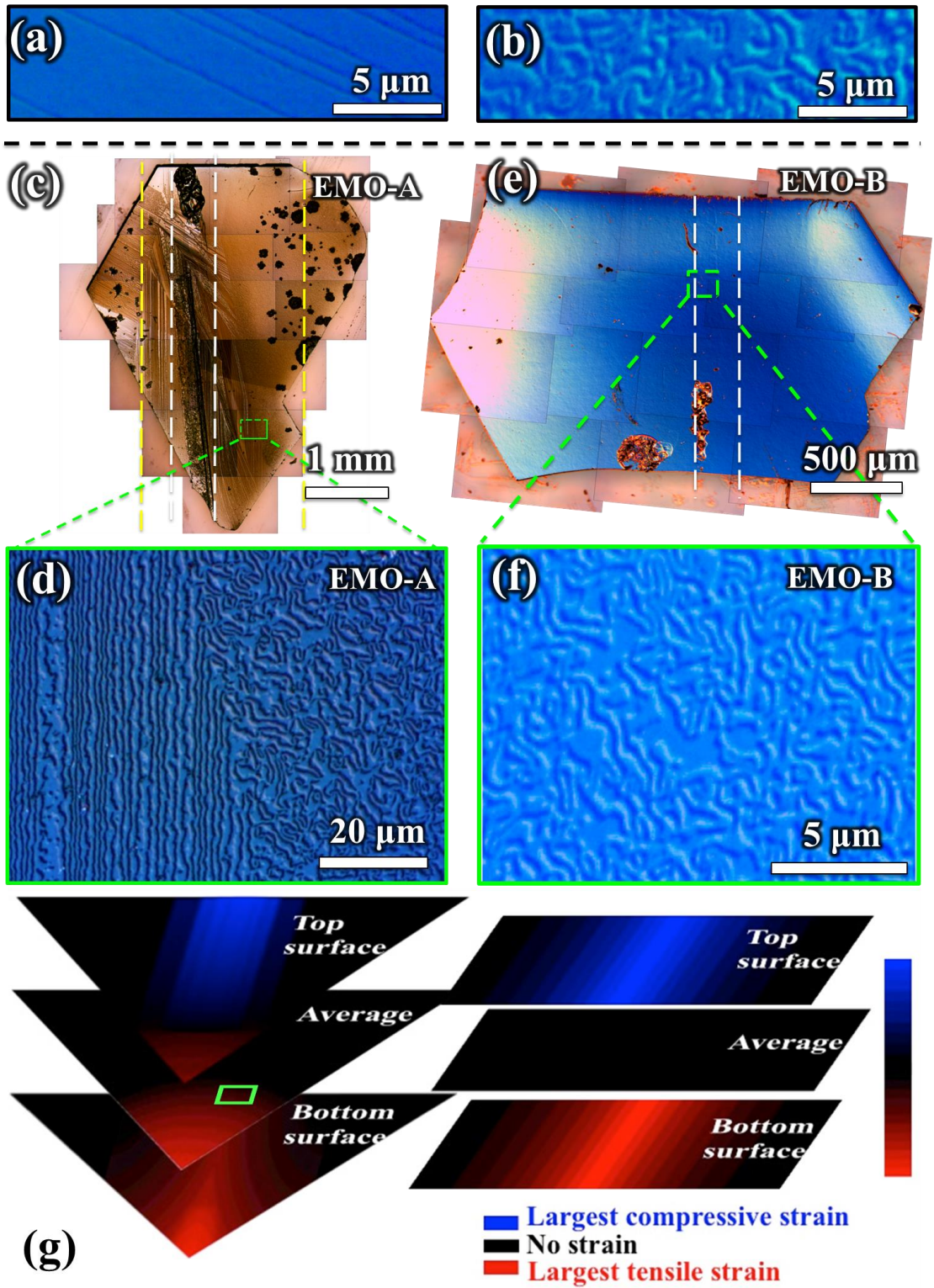

**Figure 1**

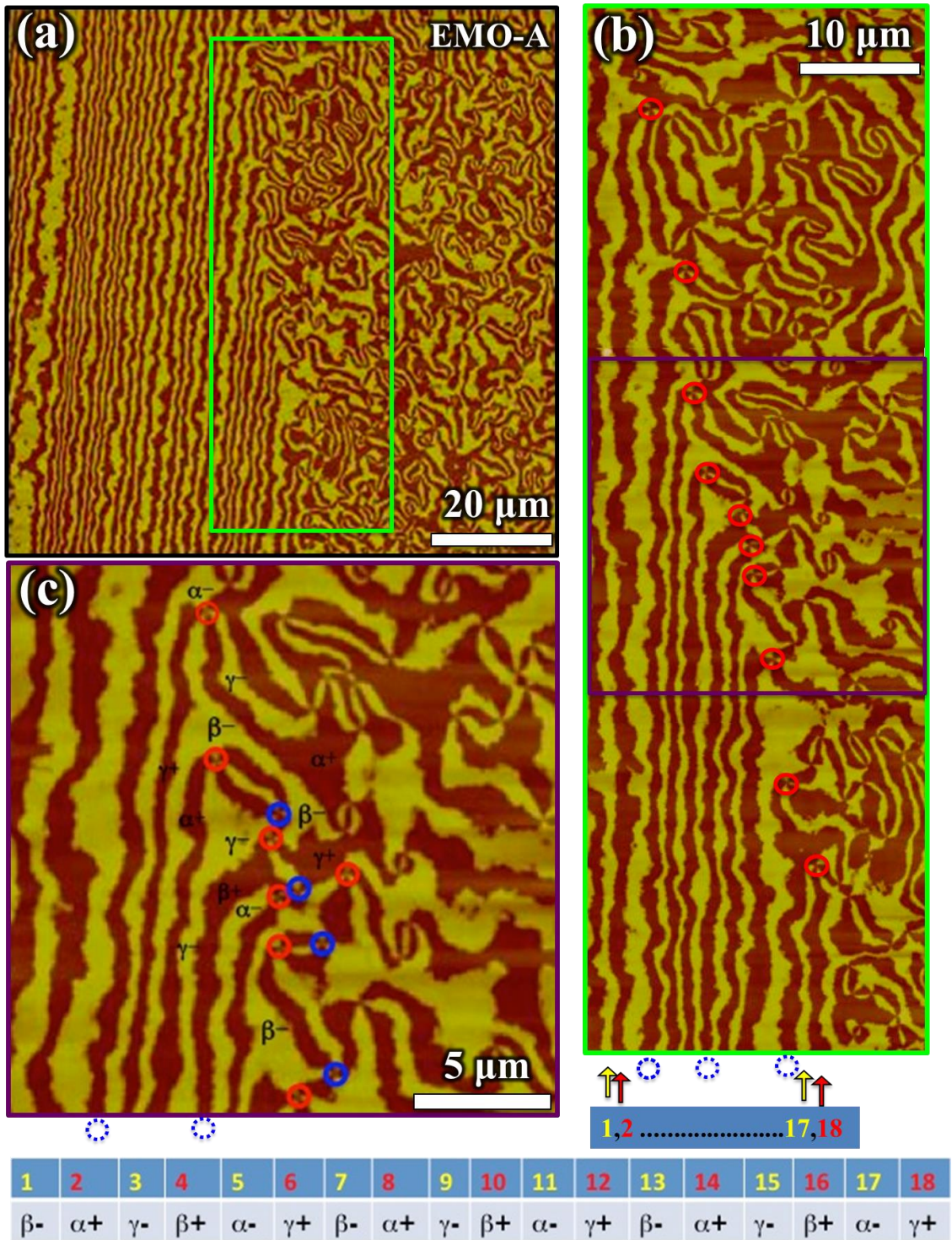

Figure 2

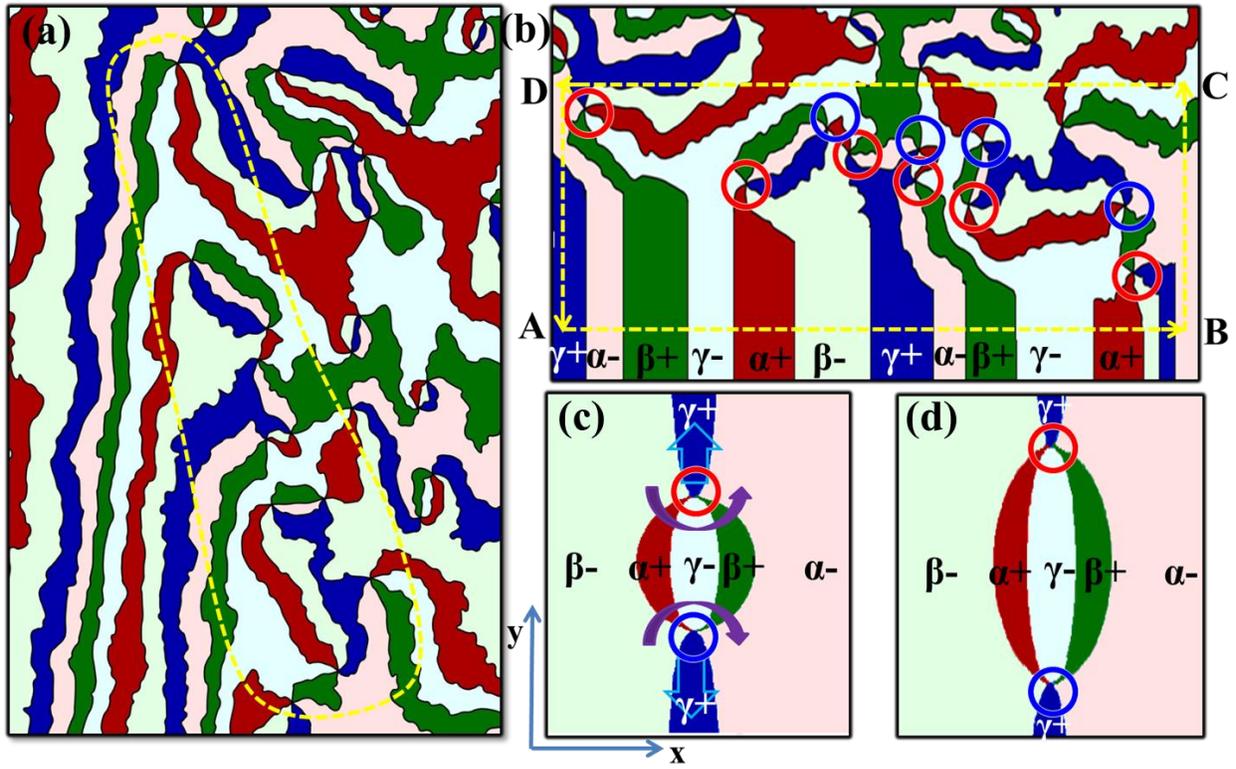

**Figure 3**

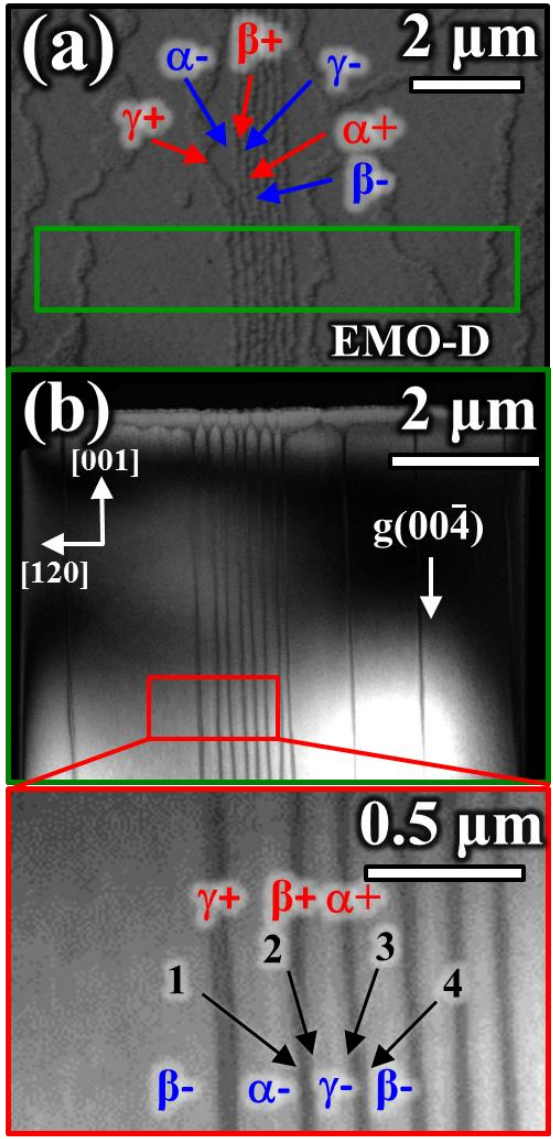
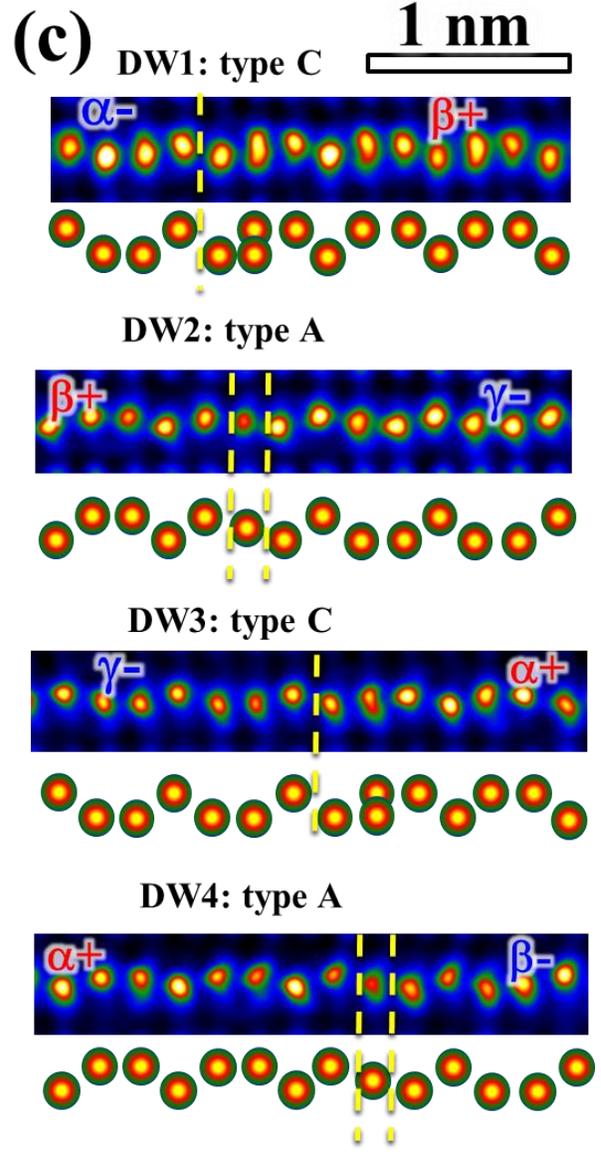

**Figure 4**

# Unfolding of vortices into topological stripes in a multiferroic material

X. Wang[1], M. Mostovoy[2], M. G. Han[3], Y. Horibe[1], T. Aoki[4], Y. Zhu[3], and S.-W. Cheong[1,*]

[1] Rutgers Center for Emergent Materials and Department of Physics and Astronomy, Rutgers University, Piscataway, NJ 08854, USA

[2] Zernile Institute for Advanced Materials, University of Groningen, Nijenborgh 4, 9747 AG, Groningen, The Netherlands

[3] Condensed Matter Physics and Materials Science, Brookhaven National Laboratory, Upton, NY 11973, USA

[4] JEOL USA, Inc., Peabody, MA 01960, USA

*Corresponding author: sangc@physics.rutgers.edu

**Supplementary Information**

**1. Applying shear strain with a large gradient to plate-like crystals of h-ErMnO$_3$**

A schematic cartoon of our strain experiment is shown in the left panel of Fig. S1. A plate-like crystal of h-ErMnO$_3$ was placed across a groove in an alumina plate. An alumina rod was put on the crystal to bend the crystal and provide a downward force. For triangular-shape crystals (see Fig. 1(c) and Fig. S3(a)), the alumina rod was slightly off-centered in the groove direction in a way that there was a larger downward force in the triangular-corner region than in the top-flat region. The right panel of Fig. S1 shows the real picture of our strain experimental setup. Note that the alumina rod is placed to be perpendicular to the top-flat edge of plate-like h-ErMnO$_3$ crystals (see Fig. 1(c), (e) and Fig. S3(a)).

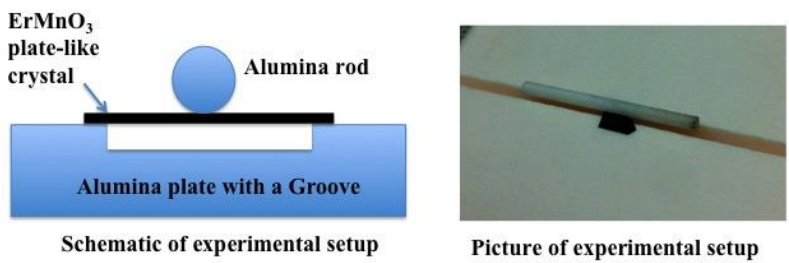

**Figure S1 |** Strain experimental setup to apply shear strain: The left schematic cartoon depicts a side view of our strain experimental setup. The right panel shows the real picture of our strain experimental setup.

## 2. Optical Microscopy images and Atomic Force Microscope (AFM) image of the vortex-to-stripe transformation areas on both surfaces of EMO-A

Fig. S2-1(a) and (b) are optical microscope images of both surfaces of plate-like EMO-A after applying strain with our strain experimental setup at high temperatures and chemical etching (see the main text for details). White dashed lines indicate the position of an alumina rod that exerts a downward force on EMO-A, and the yellow dashed lines show the location of the edges of a groove in an alumina plate. Stripe-type domains along the alumina rod direction were observed near the alumina rod location on both surfaces, and the rest area showed vortex-type domains. Fig. S2-1(c) and (d) are high-magnification optical microscope images of the green-box area in Fig. S2-1(a) and (b), respectively, showing the vortex-to-stripe transformation on both surfaces. Fig. S2-1(e) and (f) are high-magnification optical microscope images of stripe domains in the purple-box area in Fig. S2(a) and (b), respectively.

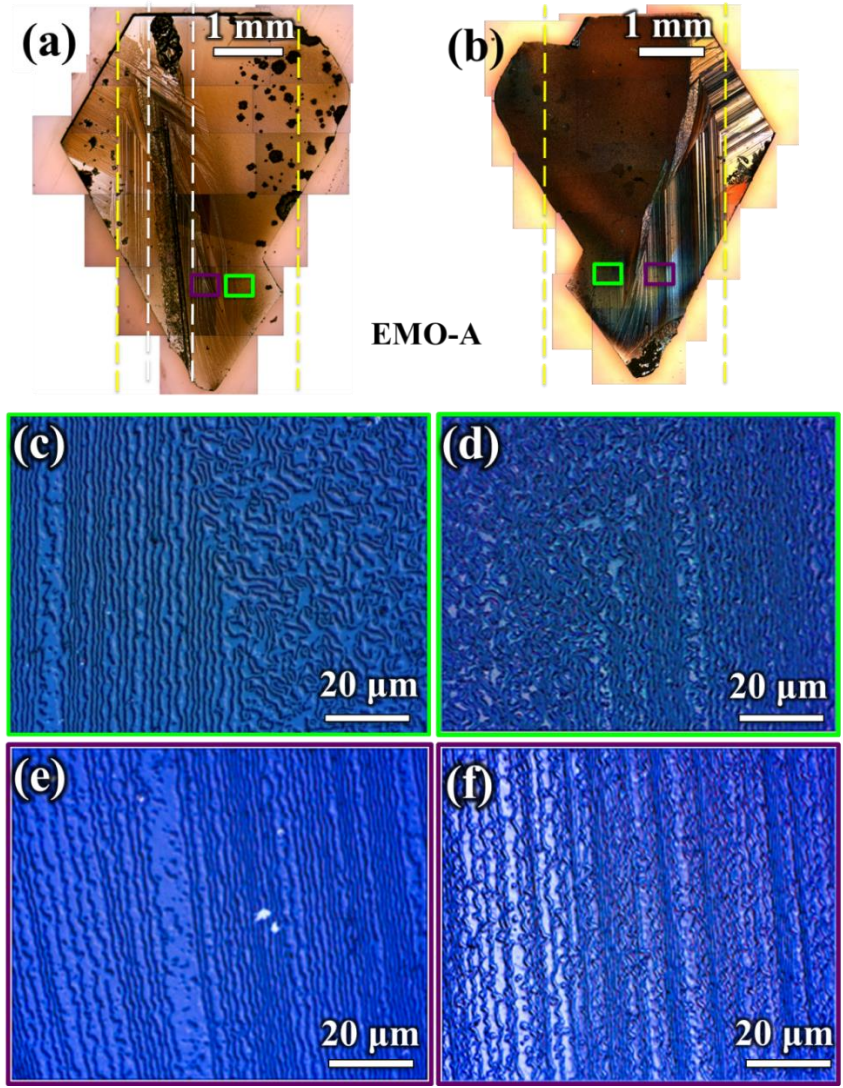

**Figure S2-1** | Optical microscope images of EMO-A after applying strain at high temperatures and chemical etching: (a) and (b) are low-magnification optical microscope images of both surfaces of EMO-A. (c) and (d) are high-magnification optical microscope images of the green-box area in Fig.S2-1(a) and (b), respectively, which show a vortex-to-stripe transformation on both surfaces. (e) and (f) are high-magnification optical microscope images of stripe domains in the purple-box region in Fig. S2-1(a) and (b), respectively.

The vortex-to-stripe transformation on the bottom surfaces of EMO-A is further studied with a AFM. Fig. S2-2 displays collaged four AFM images of a part of the area in Fig. S2-1(d), showing the vortex-to-stripe transformation. Self-consistent analysis for trimerization-ferroelectric phases of all domains indicates that all stripes exhibit monochirality, which forms through selective expulsion of antivortices (dashed blue-circles).

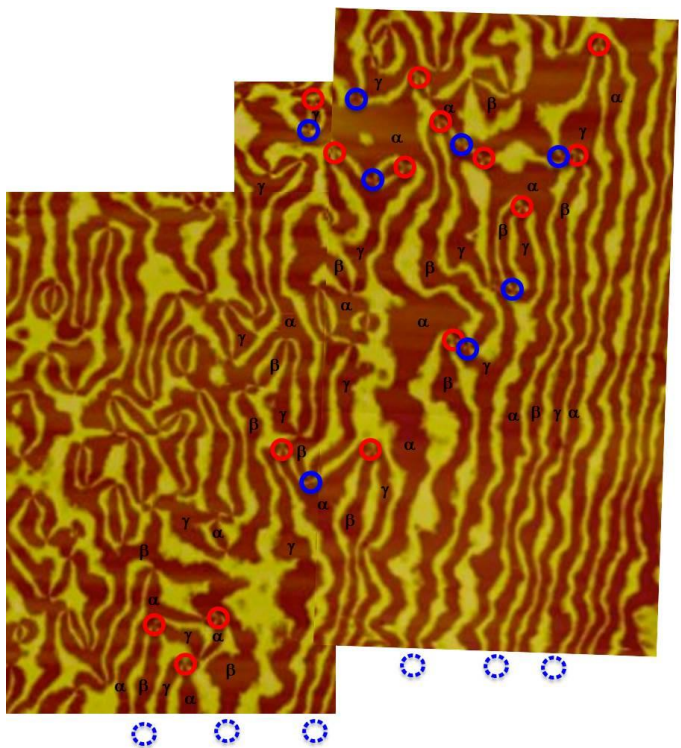

**Figure S2-2 |** AFM image of the vortex-to-stripe transformation on the second surface of EMO-A. Red circles are vortices, blue circles are antivortices, and dashed blue circles represent ejected antivortices. Vortices (red circles with the clockwise phase sequence (α-, β+, γ-, α+, β-, γ+)) exist predominantly at the boundary of a vortex-to-stripe transformation. Dashed blue circles are the expelled antivortices with the anti-clockwise phase sequence (α-, β+, γ-, α+, β-, γ+)). Stripe domains exhibit monochirality.

### 3. Optical images of another triangular-shape ErMnO$_3$ (EMO-C)

Triangular-shape crystal ErMnO$_3$ (EMO-C) clearly shows the transformation of vortices into stripes: the pink dashed line indicates the boundary between the vortex and stripe regions, which is very similar with what was observed in EMO-A. Fig. S3(b) is an enlarged optical microscope image of the green-boxed area in Fig. S3(a)

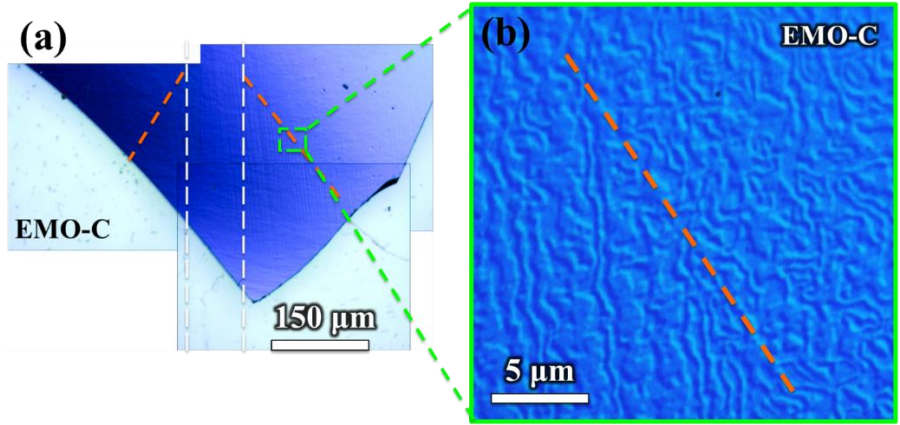

**Figure S3** | (a) Optical microscope image of triangular EMO-C. (b) High-magnification optical microscope image of the green-box area in Fig. S3(a), which shows the vortex-to-stripe transformation.

## 4. Semi-quantitative estimation of strain magnitude

Here we estimate the magnitude of strains in our samples. The values of the elastic moduli, $C_{11}$ = 18.5, $C_{33}$ = 29.8, $C_{44}$ = 9.86 and $C_{66}$ = 5.94 in units of $10^{10}$ N/m$^2$ are taken from ultrasound measurements on YMnO$_3$ (ref. [18] in main-text). The value of $C_{13}$, which does not enter into expressions for the sound velocity can be calculated using the expression for the bulk modulus,

$$B = \frac{C_{33}(C_{11} + C_{22}) - 2C_{13}^2}{C_{11} + C_{12} + 2C_{33} - 4C_{13}}, \qquad (1)$$

which for ErMnO$_3$ is 16.8 in the same units (ref. [19] in main-text). However, the maximal value of $B$ given by Eq.(1) is only about 12.5. It is reached at $C_{13} \approx 13$, which we use for estimates below.

We first consider a rectangular sample, in which case the vertical displacement of the plate, $\zeta$, only depends on the coordinate $x$ along the direction perpendicular to the cylinder axis. The boundary condition $\sigma_{\alpha\beta} n_\beta = 0$ ($\alpha,\beta = x$ or $y$), where $\mathbf{n}$ is the vector normal to the edge, gives $\sigma_{xx} = \sigma_{xy} = 0$ at the $x$-edge and $\sigma_{yy} = \sigma_{xy} = 0$ at the $y$-edge, which together with

$$\frac{\partial \sigma_{\alpha\beta}}{\partial x_\beta} = 0 \qquad (2)$$

implies the absence of average stress and strain in rectangular samples. In such samples we do not observe stripe domains.

It is also instructive to estimate the magnitude of the vertical displacement in the rectangular sample, since the form of the displacement, $\zeta(x)$, can be found exactly:

$$\zeta(x) = \zeta_0 + a\left(\frac{|x|^3}{6} - \frac{dx^2}{2}\right), \qquad (3)$$

which satisfies $\dfrac{d^4\zeta}{dx^4} = 0$ in the unloaded region and $\dfrac{d^2\zeta}{dx^2}(\pm d) = 0$ at the support line. Minimizing the sum of the potential energy of the rod and the bending energy of the plate, and using $\zeta(\pm d) = 0$, we obtain

$$\zeta_0 = -\frac{2Mgd^3}{\left(C_{11} - \dfrac{C_{13}^2}{C_{33}}\right)h^3 L}, \qquad (4)$$

where $M = 0.2$ gram is the mass of the rod, $L = 4$ cm is its length, $h = 0.02$ mm is the sample thickness and $g$ is the gravitational acceleration. For $d = 1$ mm, $\zeta_0 \approx -10^{-4}$ mm.

Furthermore, from $u_{xx} = \dfrac{du_x}{dx} + \dfrac{1}{2}\left(\dfrac{d\zeta}{dx}\right)^2 = 0,$ we find $\dfrac{du_x}{dx} = -\dfrac{9(\zeta_0 x)^2}{8d^6}(2d-|x|)^2$, i.e. the average stretching of the plate due to its vertical displacement is compensated by the in-plane compression.

While the average strain and stress for a rectangular sample are zero, the bending results in the $z$-dependent strain $u_{xx} = -z\dfrac{d^2\zeta}{dx^2} = z(d-x)\dfrac{3|\zeta_0|}{d^3}$, which is compressive on the top surface of the sample and tensile on its bottom surface ($u_{yy} = u_{xy} = 0$). This strain distribution is shown in Fig. 1(g) (right panel). Note that since $|\zeta_0| \ll h$, the bending energy in a triangular sample is much larger than the stretching energy, which is why the $z$-dependent strain in triangles is not very different from that in rectangles.

For a triangular sample $\zeta$ depends both on $x$ and $y$ and, moreover, equations for $\zeta$ and $\sigma_{\alpha\beta}$ become coupled and the boundary conditions are very complicated, which makes it impossible to find the strain distribution analytically. Therefore, we resort to dimensional analysis and order of magnitude estimates.

We first note that the second equation of the Föppl-von Karman system of equations (L. D. Landau and E. M. Lifshitz, Theory of Elasticity (Pergamon Press, 1970)) can be written in the form

$$\Delta^2 \chi + EK = 0, \qquad (5)$$

where $E = \dfrac{1}{s_{11}} = \dfrac{(C_{11}-C_{12})(C_{11}C_{33}+C_{12}C_{33}-2C_{13}^2)}{(C_{11}C_{33}-C_{13}^2)}$ and $K = \dfrac{\partial^2\zeta}{\partial x^2}\dfrac{\partial^2\zeta}{\partial y^2} - \left(\dfrac{\partial^2\zeta}{\partial x \partial y}\right)^2$ is the Gaussian curvature of the deformed sample for $\left|\dfrac{\partial\zeta}{\partial x}\right|, \left|\dfrac{\partial\zeta}{\partial y}\right| \ll 1$. If the angles in the triangle are neither small nor

close to π, the only length scale of variation of $K$ is the distance from the vertex $r$: $K \sim \frac{\zeta_0^2}{r^4}$, where $\zeta_0$ is given by Eq.(4). The stress tensor, $\sigma_{\alpha\beta} = \varepsilon_{\alpha\gamma}\varepsilon_{\beta\delta}\frac{\partial^2 \chi}{\partial x_\alpha \partial x_\beta}$, where $\varepsilon_{\alpha\beta}$ is the two-dimensional antisymmetric Levy-Civita tensor, scales then as

$$\sigma_{\alpha\beta} \sim E \frac{\zeta_0^2}{r^2}. \qquad (6)$$

The average strain distribution schematically shown in Fig. 1(g) (left panel) is consistent with this scaling. We note that there is no stress singularity at $r = 0$, as Eq.(6) is not applicable in the region where the rod touches the sample.

**5. Proof of the forces on vortices and antivortices resulting from Lifshitz term**

Consider a single domain wall connecting a vortex with an antivortex, as shown in Fig. S5-1 (clearly, a domain wall cannot connect two vortices or two antivortices). At the domain wall the trimerization phase $\Phi$ increases by π/3 and we assume that $\Phi = 0$ on the right hand side of the wall beginning at the vortex and ending at the antivortex and $\Phi = \pi/3$ on the left hand side. Furthermore, we introduce local Cartesian coordinates $(\xi, \eta)$, such as $\xi$ varies along the domain wall and $\eta$ in the direction perpendicular to the wall, so that

$$\frac{\partial \Phi}{\partial \eta} = \frac{\pi}{3}\delta(\eta).$$

Then the Lifshitz invariant (Eq.(1) in the main text) can be written as

$$F_{int} = -\lambda h \int \frac{d\xi d\eta}{|\det J|} [(u_{xx} - u_{yy})\partial_x \eta - 2u_{xy}\partial_y \eta]\delta(\eta)$$

$$= -\lambda h \frac{\pi}{3} \int d\xi \left[(u_{xx} - u_{yy})\frac{\partial_x \eta}{\det J} - 2u_{xy}\frac{\partial_y \eta}{\det J}\right].$$

Here, $h$ is the sample thickness and $J$ is the transformation matrix:

$$J = \begin{bmatrix} \partial_x \xi & \partial_y \xi \\ \partial_x \eta & \partial_y \eta \end{bmatrix}.$$

The directions of the $\eta$ and $\xi$ axes are chosen such that the determinant of $J$ is positive (see Fig. S5-1), so that $|\det J| = \det J$.

Using the expression for the inverse transformation matrix,

$$J^{-1} = \begin{bmatrix} \partial_\xi x & \partial_\eta x \\ \partial_\xi y & \partial_\eta y \end{bmatrix} = \frac{1}{\det J}\begin{bmatrix} \partial_y \eta & -\partial_y \xi \\ -\partial_x \eta & \partial_x \xi \end{bmatrix},$$

We can then re-write $F_{int}$ for the uniform strain in the form,

$$F_{int} = \lambda h \frac{\pi}{3} \int d\xi [(u_{xx} - u_{yy})\partial_\xi y + 2u_{xy}\partial_\xi x]$$

$$= \lambda h \frac{\pi}{3}[(u_{xx} - u_{yy})(y_A - y_V) + 2u_{xy}(x_A - x_V)],$$

where $(x_V, y_V)$ and $(x_A, y_A)$ are the coordinates of the vortex and antivortex, respectively. The forces acting on vortex and antivortex are then given by

$$\mathbf{f}_V = -\frac{\partial F_{int}}{\partial \mathbf{x}_V} = -\mathbf{f}_A = \lambda h \frac{\pi}{3}\left(2u_{xy}, (u_{xx} - u_{yy})\right).$$

The result of the action of these forces is illustrated in Fig. S5-2. If, for example, $u_{xy} = 0$ and $\lambda(u_{xx} - u_{yy}) > 0$ the Magnus-type force acting on the vortex is applied along the $y$ axis (Fig. S5-2a). It favors vertical stripes, as shown in Fig. S5-2b. If, on the other hand, $(u_{xx} - u_{yy}) = 0$ and $\lambda u_{xy} > 0$, the force is applied the $x$ axis (Fig. S5-2c), resulting in horizontal stripes (Fig. S5-2d). In both cases the sign of the phase $\Phi$ jump across the wall is such that the interaction energy $F_{int}$ is minimal.

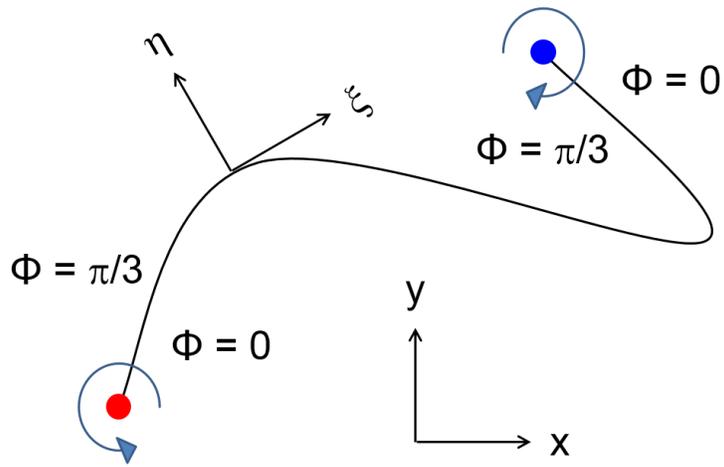

**Fig. S5-1** A pair of vortex and antivortex connected by a single domain wall.

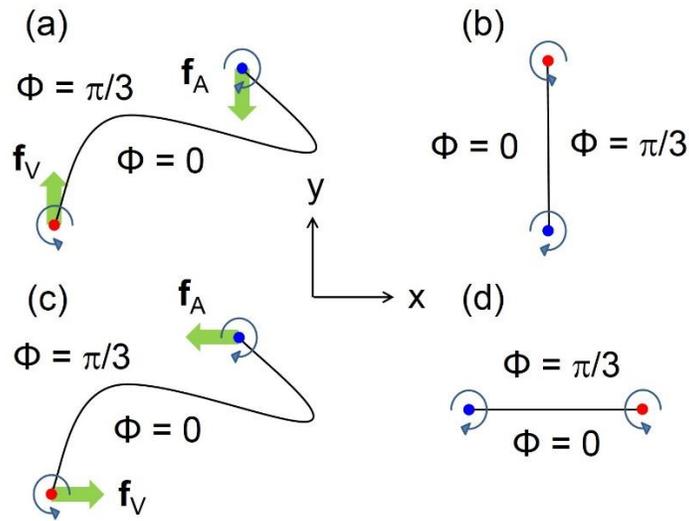

**Fig. S5-2** Magnus-type forces on vortex and antivortex: (a) and (b) shows when $u_{xy} = 0$ and $\lambda(u_{xx} - u_{yy}) > 0$, the force acting on the vortex is along the $y$ axis, resulting in vertical stripes are favored. (c) and (d) shows when $(u_{xx} - u_{yy}) = 0$ and $\lambda u_{xy} > 0$, the force is along the $x$ axis, giving the horizontal stripes.

## 6. Forces on vortices induced by strain and strain gradient

Fig. S6-1 explains pictorially how a shear strain coupled to the phase gradient by the interaction $\int dV \left[ (u_{xx} - u_{yy}) \partial_x \Phi - 2 u_{xy} \partial_y \Phi \right]$ can lead to stripes with monochirality. If we assume that the order of ($\gamma$-, $\beta$+, $\alpha$-, $\gamma$+, $\beta$-, $\alpha$+ from left to right) is favored by the term $u_{xx} \partial_x \Phi$ in the presence of a horizontal (the *x*-direction) shear strain, then the domain walls following the purple arrows 2 and 3 have a lower energy then the domain walls following yellow arrows 1 and 4. Then, the upward motion of a vortex and downward motion of antivortex will enlarge the region with lower energy walls and shrink the region with higher energy walls, which will lead to lowering of the total energy. This effect will result in strain-induced Magnus-type force acting on vortices and antivortices in opposite directions.

Therefore, the strain favors the topological stripe domain state with a nonzero average gradient of the phase $\Phi$, which forces vortex and antivortex to move in opposite directions (normal to the shear strain direction). This state is an analog of the current-carrying states of superconductors and superfluids, where the supercurrent $\boldsymbol{j} \propto \nabla \Phi$. The Magnus force in such states pulls vortices away from antivortices in the direction normal to $\boldsymbol{j}$ (or that of the average phase gradient). This effect is similar to the transverse force in h-RMnO$_3$, the difference being the smooth variation of the condensate phase vs discontinuous jumps of the phase of the periodic lattice modulation at domain walls in h-RMnO$_3$s.

The domain patterns observed in EMO-A and EMO-C require additional considerations. If there exists only the shear strain effect discussed above, then one would expect to find a huge density of "vortices" at the vortex-stripe transformation boundary, because all vortices that originally occupied the large stripe area should accumulate at the transformation boundary, while all corresponding antivortices should be removed from the specimen. Since we observe only a small excess of vortices at the transformation boundary proportional to the number of stripes (see main text), we propose an addition mechanism: both vortices and antivortices can be moved by the strain gradient in the same direction, which would explain how many vortex-antivortex pairs can be expelled from the specimen.

This force induced by strain gradient can results from expansion/contraction of lattice around vortices described, for example, by the term,

$$F_{\text{int}}^{(2)} = g(u_{xx} + u_{yy})(\partial_i Q \partial_i Q + Q^2 \partial_i \Phi \partial_i \Phi)$$

where $Q$ is the amplitude of the order parameter (lattice trimerization) and $\Phi$ is its phase. For an inhomogeneous strain $u_{xx} + u_{yy}$ depending on the $(x,y)$ coordinates in the $ab$-plane, this coupling is effectively a potential energy, $U(x,y)$, of the vortex which, depending on the sign of the coupling constant $g$, forces the vortex to move into regions with larger or smaller strain. Since $F_{int}^{(2)}$ is independent of sign of the phase gradient, it gives rise to a force that moves vortex and antivortex in the same direction. The result of the two different couplings to strains is the formation of monochiral stripes with "only a small excess amount of vortices" remaining at the transformation boundary. We note that the interaction $F_{int}^{(2)}$ also leads to strain-mediated interactions between vortices, as was discussed in the context of vortex arrays in superconductors (refs. [22-23] in main-text).

We also note that for the bending deformation caused by the alumina rod, the non-zero components of the strain tensor, $u_{xx}$, $u_{yy}$, $u_{xy}$ and $u_{zz}$, are proportional to the z-coordinate (normal to the surface) and have opposite signs at the upper and lower surfaces of the sample [see, for example, Landau, L. D. & Lifshitz, E. M. Theory of Elasticity (Pergamon, Oxford, 1986), page 44]. Such a strain favors stripes with opposite signs of the gradient of the phase $\Phi$ at the two surfaces, which are energetically very costly because a nonzero $\oint d\mathbf{x} \cdot \nabla \Phi$ around the sample surface implies the presence of an array of parallel vortex cores inside the sample. This explains why we do not observe stripes in EMO-B.

Both topological and non-topological couplings to strain originate from lattice anharmonicity and are proportional to the second power of the trimerization amplitude $Q$. The energy of the topological interaction per unit area of the domain wall can be estimated as $u\frac{Q^2}{a^4}Ry$, where $u$ is the strain, $a$ is the lattice constant and Ry is Rydberg constant. Similarly, the energy of the non-topological interaction per unit length of the vortex line is $\sim u\frac{Q^2}{a^3}Ry$. Thus for a vortex displaced by a length $L$, the ratio of the non-topological and topological interactions is $\sim \frac{a}{L}\frac{(u(L)-u(0))}{u} \ll 1$, since the typical width of the stripe phase is much larger than one lattice constant. Although weak, the coupling to strain gradient can play an important role for kinetics of domain wall and vortices near the phase transition, e.g. by moving crystallographic defects accumulated at vortices and domain walls and hindering their motion.

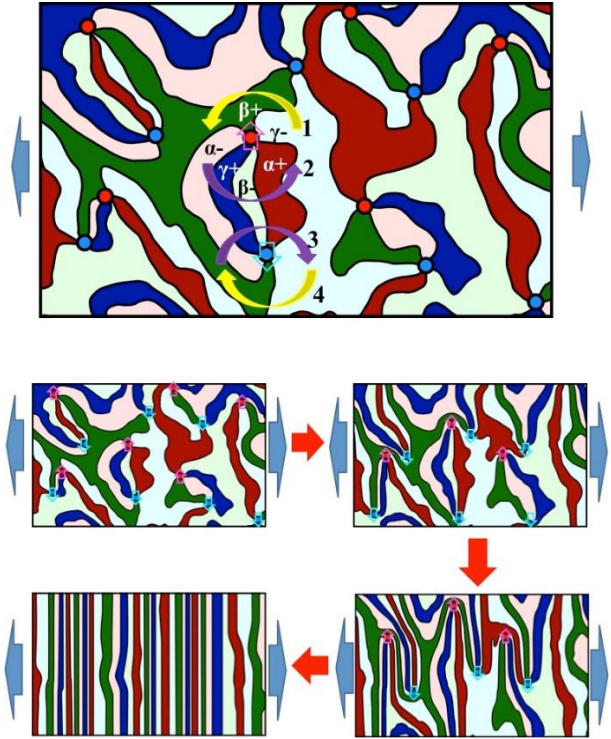

**Figure S6-1 |** Magnus-type force pushing vortices and antivortices in opposite direction. Schematics of the unfolding process of vortices/antivortices to mono-chiral stripes induced by horizontal shear strain (large blue arrows). The shear strain induces a Magnus-type force (open arrows) acting on vortices (red) and antivortices (sky-blue) in opposite directions.

Fig. S6-2 (a) displays a cartoon showing an exaggerated displacement of a triangular-shape crystal under an off-centered alumina rod. The in-plane strain distribution on a triangular-shape crystal in our strain experimental setup is schematically shown in Fig. S6-2(b). The top compressive strain and the bottom tensile strain are canceled to each other in average or in the middle of the crystal. The only in-plane strain left over in average is shear strain and locates near the triangular corner of the crystal.

We found that stripe domains walls have a strong tendency to orient along the [110] directions (i.e. perpendicular to the top crystal edge in Fig. 1(c) and Fig. S3 (a)). This effect, combined with the Magnus-type force due to the shear strain along the *x* direction and the strain gradient along the *y* direction results in the preferential expulsion of antivortices and the concurrent formation of topological stripes that orient along the direction perpendicular to the top flat edge of EMO-A or EMO-C. There are no vortices/antivortices left in the stripe region. A relatively small amount of unfolded vortices is accumulated at the boundary between strained region and not-strained regions.

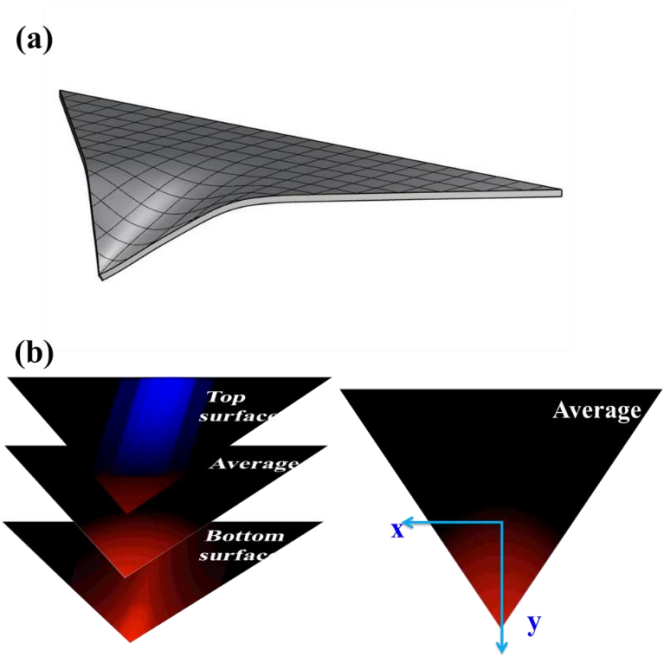

**Figure S6-2** | Schematic illustration of in-plane strain distribution with a large gradient for a triangular-shape crystal. The top panel shows a perspective view of the bending of a triangular-shape crystal under an alumina rod at high temperatures. In-plane strain contour plots on the top surface and bottom surface, and in average are shown in the left panel. Blue, red, and black colors indicate compressive, tensile, and no strain. The average shear strain in the right panel is along the *x* direction, and the gradient of the average shear strain along the *y* direction is positive.

Fig. S6-3 shows schematic cartoons for a "simple" situation of having only isolated vortex and antivortex pairs. We start with a random distribution of isolated vortex-antivortex pairs - all colors and labels are identical with what we use in the main text, and then we compare 3 different cases: (1) only Magnus-type force without strain-gradient-induced force, (2) only strain-gradient-induced force without Magnus-type force, and (3) Magnus-type force plus strain-gradient-induced force (which is consistent with our experimental observation). From these cartoons, it is clear that if we have only the Magnus-type force, then there should be a large excess of vortices at the vortex-stripe transition boundary, because only antivortices are expelled from the sample, while the relevant vortices accumulate at the transition boundary, and the number of vortices remains the same. If we also take the strain-gradient-induced force into account, then we obtain topological stripes with a smaller number of vortices at the vortex-stripe transition boundary, as is observed experimentally. Note that if we consider only the strain-gradient-induced force (case (2) in Fig. S6-3), then the trimerization phases in the stripes do NOT follow the sequence of (α+, β-, γ+, α-, β+, γ-), which is inconsistent with our experimental observation. Therefore, both the Magnus-type and strain-gradient-induced forces are necessary to explain our experimental observation, resulting topological stripes.

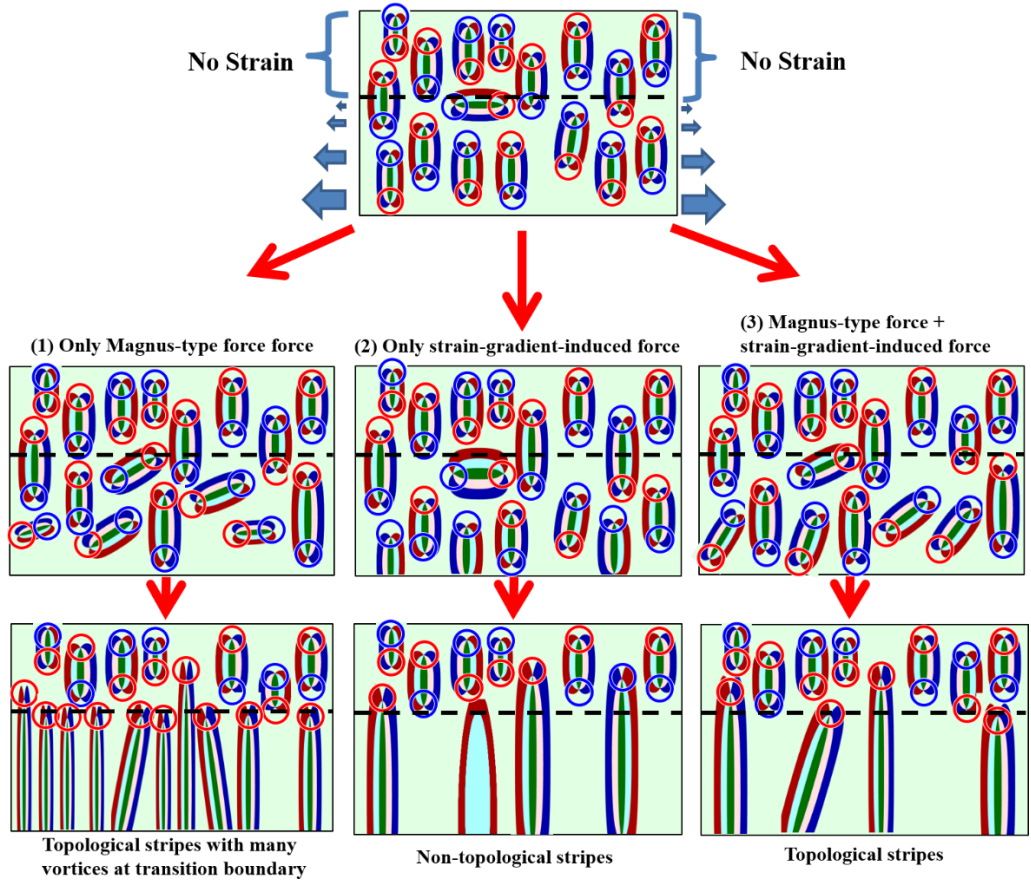

**Fig. S6-3** Top: beginning stage of a simple situation with randomly-distributed isolated vortex-antivortex pairs. Above the black dashed line, there is no strain, while below the black dashed line, there exists a non-uniform shear strain (blue arrows), which reflects our experimental situation with triangular crystals. We consider three distinct cases: (1) If only the Magnus-type force is present, topological stripes with a high density of vortices at the vortex-stripe transition boundary are formed, which is inconsistent with our experimental observation (left-bottom two cartoons). (2) If only the strain-gradient-induced force is present, the resulting stripes do not follow the sequence of (α+, β-, γ+, α-, β+, γ-) and thus are non-topological (middle-bottom two cartoons). (3) If both Magnus-type force and strain-gradient-induced forces are present, the topological stripes with a smaller number of vortices near the stripe domain boundary are formed, which is consistent with our experimental observation (right-bottom two cartoons).

## 7. Atomic structures of topological stripe domain walls

Fig. S7(a) to (d) show HAADF-STEM images of domain walls in Fig. 4 in the main text. All images are deconvoluted using a maximum entropy method (HREM Research Inc.). The areas used for Fig. 4 are indicated with white rectangular boxes. Fig. S7(e) displays the line profiles from the area indicated with red arrows in Fig. S7(a) to (d). It is apparent that Er columns at the C type walls (DW1 and DW3) consist of two separate peaks, displaced upward ($Er_{up}$) or downward ($Er_{down}$). The separation of the peaks is about 0.05 nm. On the other hand, Er columns at the A type walls (DW2 and DW4) exhibit only a single peak, located in the middle of upward and downward Mn columns.

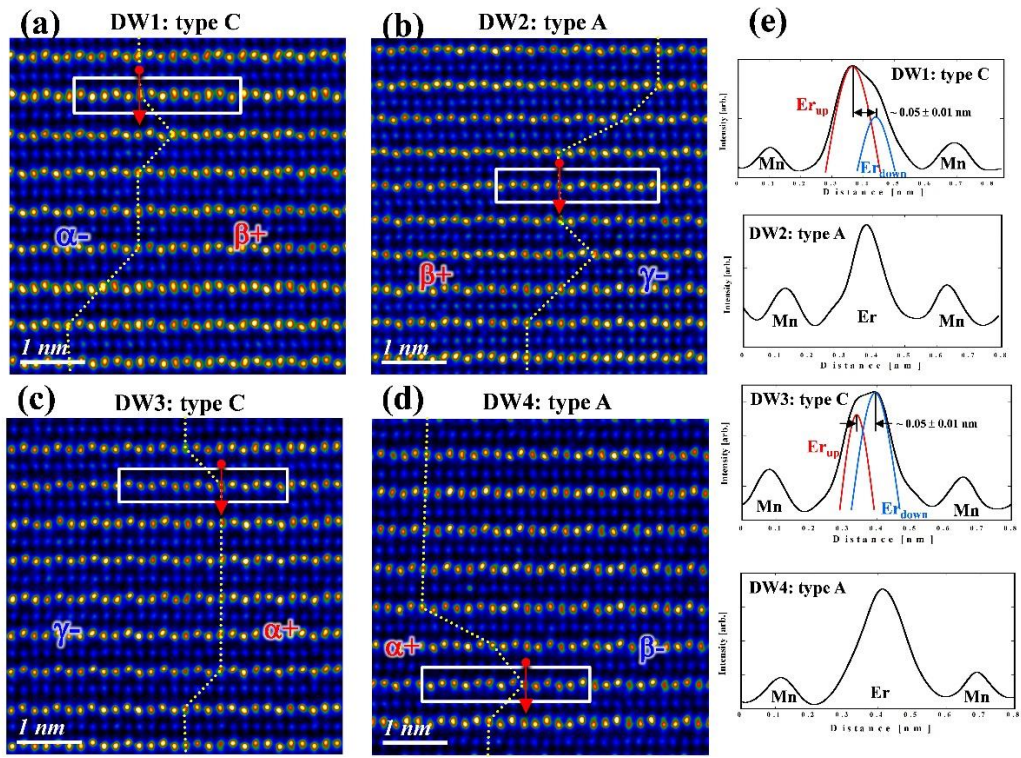

**Figure S7** | HAADF-STEM images of topological stripe domain walls. (a) to (d) are full HAADF-STEM images of domain walls shown in Fig. 4(b). White rectangular boxes indicate the area shown in Fig. 4(c). (e) shows the line profiles for the width indicated with red arrows in a to d.

## 8. Local ionic distortions at four different types of domain walls

The crystallographic structures of 4 different types (A-D) of domain walls (DW) are shown in Fig. S8. Yellow, brown, and light-blue circles represent Er, Mn and O ions, respectively. Light blue and dark blue circles indicate the top and bottom apical oxygen ions of MnO$_5$ hexahedra, respectively. Arrows show the directions of atomic displacements. The triangles with green bars correspond to Mn trimers. Blue dashed lines indicate the domain walls. A type and B type are walls with un-distorted Er-ions, and C type and D type are walls with distorted Er-ions. There exists 2π/3 antiphase shift at A type and C type, while B type and D type shows 4π/3 antiphase shift.

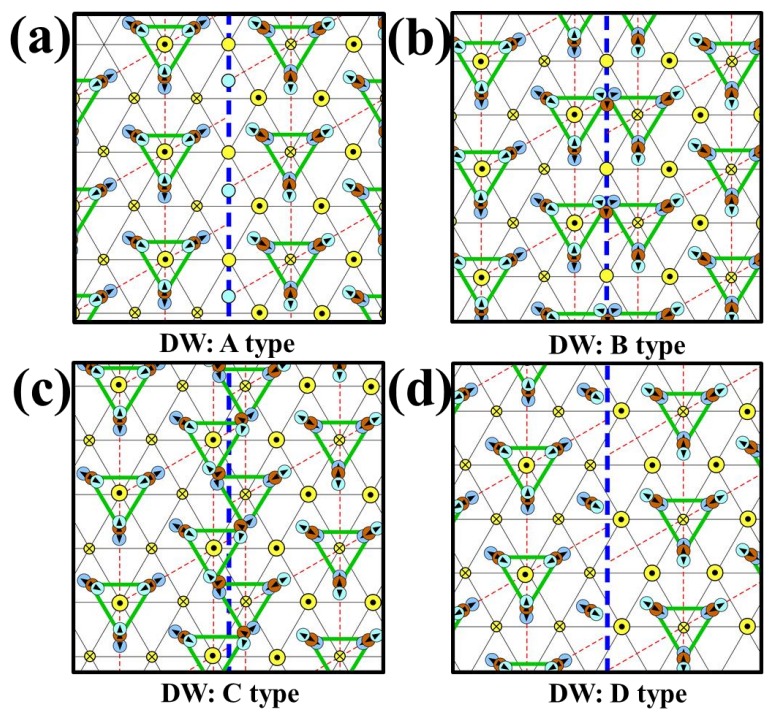

**Figure S8** | Crystallographic structures of 4 different domain walls: Yellow, brown, and light-blue circles represent Er, Mn and O ions, respectively. Light blue and dark blue circles indicate the top and bottom apical oxygen ions of MnO$_5$ hexahedra, respectively. Arrows show the directions of atomic displacements. The triangles with green bars correspond to Mn trimers. Blue dashed lines indicate domain walls.

**9. Local ionic distortions at domain walls in a vortex-to-stripe transformation region**

Fig. S9(a) and (b) are an AFM image showing a vortex-to-stripe transformation and the corresponding schematic with colored trimerization-ferroelectric phases. Red, blue, and green colors represent three different types of trimerization antiphase (α, β, γ) and dark and light colors indicate ferroelectric polarization directions (up and down, respectively). Local ionic distortions corresponding to the yellow- and light-blue-box areas are depicted in Fig. S9(c) and (d), respectively. Blue dashed lines indicate connected domain walls from vortex to stripe domains. In this schematic, the B type domain wall between [β-, α+] near the vortex core switches to A type in the stripe area, while the C type domain wall between [α+, γ-] remains unchanged. Note that B type and C type walls exist at an angle with 60 degrees, and A type and C type walls are parallel to each other.

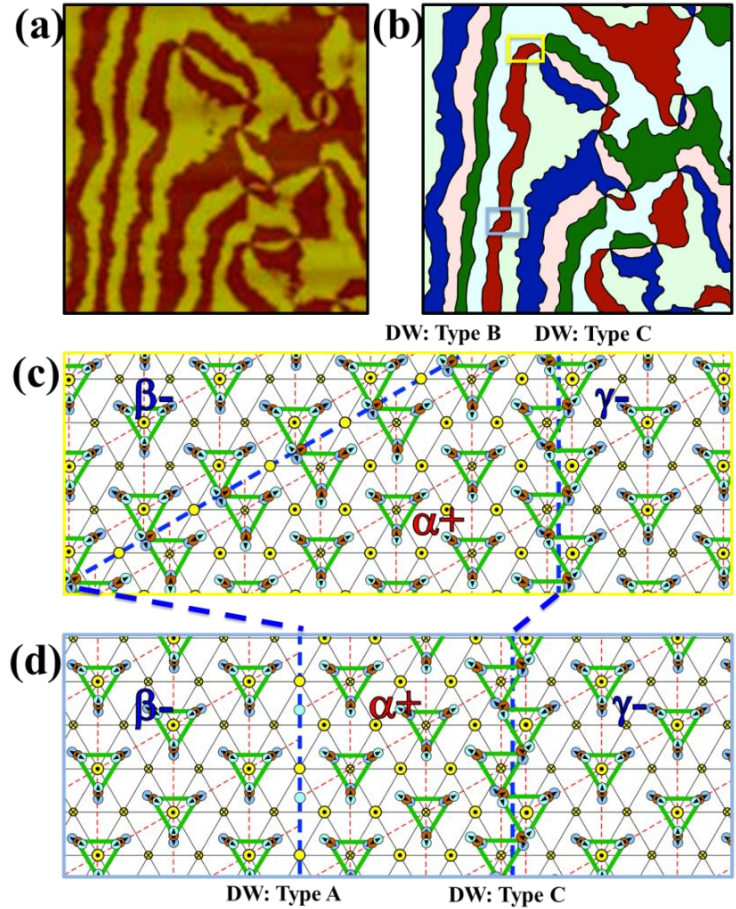

**Figure S9** | Local ionic distortions for a vortex-to-stripe transformation. (a) AFM image showing a vortex-to-stripe transformation. (b) Schematic cartoon corresponding to Fig. S9(a), with colored antiphase and
35Fig. S9(a) and (b) are an AFM image showing a vortex-to-stripe transformation and the corresponding schematic with colored trimerization-ferroelectric phases. Red, blue, and green colors represent three different types of trimerization antiphase (α, β, γ) and dark and light colors indicate ferroelectric polarization directions (up and down, respectively). Local ionic distortions corresponding to the yellow- and light-blue-box areas are depicted in Fig. S9(c) and (d), respectively. Blue dashed lines indicate connected domain walls from vortex to stripe domains. In this schematic, the B type domain wall between [β-, α+] near the vortex core switches to A type in the stripe area, while the C type domain wall between [α+, γ-] remains unchanged. Note that B type and C type walls exist at an angle with 60 degrees, and A type and C type walls are parallel to each other.

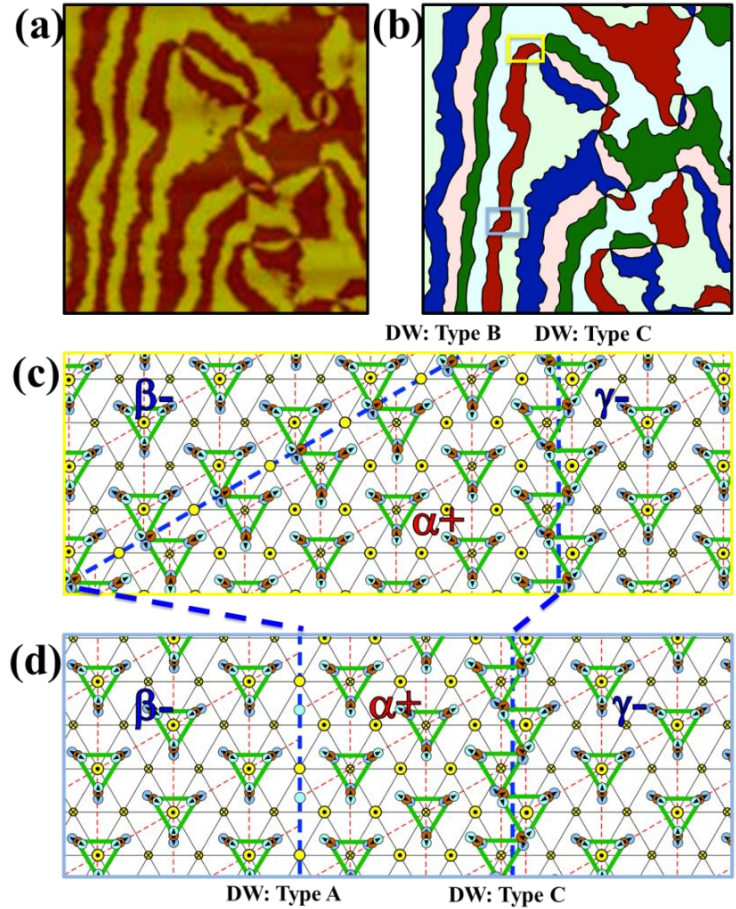

**Figure S9** | Local ionic distortions for a vortex-to-stripe transformation. (a) AFM image showing a vortex-to-stripe transformation. (b) Schematic cartoon corresponding to Fig. S9(a), with colored antiphase and

ferroelectric domains. (c) and (d) depict local ionic distortions corresponding to the yellow- and light-blue-box areas, respectively. Blue dashed lines indicate connected domain walls from vortex to stripe domains.